\documentclass[conference]{IEEEtran}
\IEEEoverridecommandlockouts
\usepackage{flushend}
\usepackage{amsmath,graphicx,cite}
\usepackage{bm}
\usepackage{amsthm, amssymb}
\usepackage[ruled]{algorithm}
\usepackage{algpseudocode}
\usepackage{algpascal}
\usepackage{algc}
\usepackage{url}
\usepackage{color}

\ifCLASSINFOpdf
\else
\fi
\newcommand{\V}[1]{\bm{#1} } %vector command

\newcommand{\mR}{\mathbb{R}}

\newcommand{\lb}{\left(}
\newcommand{\rb}{\right)}
\newcommand{\lbb}{\left\{}
\newcommand{\rbb}{\right\}}

 \newcommand{\LOOE}{\epsilon_{\rm LOO}}
 \newcommand{\LOOEtilde}{\tilde{\epsilon}_{\rm LOO}}

\newcommand{\nnzero}{K}
\newcommand{\nnzeroset}{S_{K}}
\newcommand{\nnzeromax}{K_{\rm max}}

\newcommand{\Req}[1]{(\ref{eq:#1})}

\newcommand{\Rfig}[1]{Fig.\ \ref{fig:#1}}

\newcommand{\Lfig}[1]{\label{fig:#1}}
\newcommand{\Leq}[1]{\label{eq:#1}}
\newcommand{\Rsec}[1]{sec.\ \ref{sec:#1}}
\newcommand{\Lsec}[1]{\label{sec:#1}}

\newcommand{\be}{\begin{eqnarray}}
\newcommand{\ee}{\end{eqnarray}}
\newcommand{\ba}{\begin{array}}
\newcommand{\ea}{\end{array}}

\newcommand{\subbe}{\begin{subequations}}
\newcommand{\subee}{\end{subequations}}
\newcommand{\bs}{\backslash}
\newcommand{\mc}[1]{\mathcal{#1}}
\DeclareMathOperator*{\argmin}{arg\,min}

\newcommand{\lA}{\leftarrow}

\newcommand{\Lcode}[1]{\label{code:#1}}
\newcommand{\Rcode}[1]{{ Algorithm\ \ref{code:#1}}}

\begin{document}
\title{Sampling approach to sparse approximation problem: determining degrees of freedom by simulated annealing}

\author{\IEEEauthorblockN{Tomoyuki Obuchi and Yoshiyuki Kabashima}
%\IEEEauthorblockA{Department of Mathematical and Computing Science,\\
\IEEEauthorblockA{Department of Mathematical and Computing Science,\\
 Tokyo Institute of Technology, Yokohama 226-8502, Japan}
%\and
%\IEEEauthorblockN{Yoshiyuki Kabashima}
%\IEEEauthorblockA{Interdisciplinary Graduate School of Science and Engineering,\\
% Tokyo Institute of Technology, Yokohama, Kanagawa 226-8502, Japan}
%\and
%\IEEEauthorblockN{James Kirk\\ and Montgomery Scott}
%\IEEEauthorblockA{Starfleet Academy\\
%San Francisco, California 96678--2391\\
%Fax: (888) 555--1212}
}


%Telephone: (800) 555--1212, Fax: (888) 555--1212}


\maketitle

\begin{abstract}
The approximation of a high-dimensional vector by a small combination of column vectors
selected from a fixed matrix has been actively debated in several different disciplines.
In this paper, a sampling approach based on the Monte Carlo method is presented as an
efficient solver for such problems. Especially, the use of simulated annealing (SA),
a metaheuristic optimization algorithm, for determining degrees of freedom (the number of used columns)
by cross validation is focused on and tested. 
Test on a synthetic model indicates that our SA-based approach can find a nearly optimal solution for the approximation problem and, when combined with the CV framework, it can optimize the generalization ability. Its utility is also confirmed by application to a real-world supernova data set.
\end{abstract}


% \fi
\IEEEpeerreviewmaketitle

\section{Introduction} \Lsec{Introduction}
%In the formulation of compressed sensing,
In a formulation of compressed sensing,
a sparse vector $\V{x} =(x_i) \in \mathbb{R}^N$ is recovered from
a given measurement vector $\V{y}=(y_\mu) \in \mathbb{R}^M$ $(M < N)$
by minimizing the residual sum of squares (RSS)
under a sparsity constraint as
\be
\hat{\V{x}}(K)=
\argmin_{\V{x}}
\left \{\frac{1}{2}||\V{y}-A\V{x}||_2^2 \right \}
\ \mbox{\rm  subj.  to } \ ||\V{x}||_0 \le \nnzero,
\Leq{SAP}
\ee
where
$A=(A_{\mu i}) \in \mathbb{R}^{M \times N}$ and $||\V{x}||_0$ denote
the measurement matrix and the number of non-zero components in $\V{x}$
($\ell_0$-norm), respectively \cite{Elad:10}.
The need to solve similar problems also arises in many contexts of information science
such as variable selection in linear regression, data compression, denoising, and machine learning.
We hereafter refer to the problem of \Req{SAP} as the {\em sparse approximation problem} (SAP).

Despite the simplicity of its expression, SAP is
highly nontrivial to solve. Finding the exact solution of \Req{SAP} has proved to be NP-hard \cite{Natarajan:95},
and various approximate methods have been proposed.
Orthogonal matching pursuit (OMP) \cite{Pati:93}, in which the set of used columns
is incremented in a greedy manner to minimize RSS, is
a representative example of such approximate solvers.
Another way of finding an approximate solution of \Req{SAP} is 
by converting the problem into a Lagrangian form
$\frac{1}{2}||\V{y}-A\V{x}||_2^2+\lambda ||\V{x}||_p^p$
in conjunction with relaxing $||\V{x}||_0$ to $||\V{x}||_p^p=\sum_{i=1}^N |x_i|^p$.
In particular, setting $p=1$
makes the converted problem convex and allows us to efficiently find its unique minimum solution.
This approach is often termed LASSO~\cite{Tibshirani:96}.
When prior knowledge about the generation of $\V{x}$ and $\V{y}$
is available in the form of probability distributions, one may resort to
the Bayesian framework for inferring $\V{x}$ from $\V{y}$.
This can be efficiently carried out by approximate message passing (AMP)~\cite{Krzakala:12}.

Solvers of these kinds have their own advantages and disadvantages, and
the choice of which to employ depends on the imposed constraints and available resources.
%OMP exhibits fairly good performance when $\nnzero$ is small enough, but
%the approximation accuracy deteriorates as $\nnzero$ grows larger.
%The uniqueness of solution of the $\ell_1$-regularization approach
%may be preferred for theoretically guaranteeing properties of the solution.
%However, the existence of the $\ell_1$-regularizer prevents the solution from
%truly minimizing RSS. The Bayesian framework offers excellent performance
%when the assumed prior and noise distributions are correct, which, however,
%does not necessarily hold in practical situations.
This means that developing novel possibilities is important for offering more choices.
Supposing a situation where one can make use of relatively high computational resources,
we explore the abilities and limitations of another approximate approach,
i.e., Monte Carlo (MC)-based sampling.

In an earlier study, we showed that a version of simulated annealing (SA)~\cite{Kirkpatrick:83}, which is a versatile MC-based metaheuristic for functional optimization, has the ability to efficiently find a nearly optimal solution for SAP in a wide range of system parameters~\cite{Obuchi:16a}.
In this paper, we particularly focus on the problem of determining the {\em degrees of freedom}, $\nnzero$, by cross validation (CV)
utilizing SA. We will show that the necessary computational cost of our algorithm is bounded by $O(M^2 N |\nnzeroset| \nnzeromax   )$,  where $\nnzeroset$ is the set of tested values of $K$ and $|\nnzeroset|$ is its cardinality, and $\nnzeromax$ is the largest one among tested values of $K$. Admittedly, this computational cost is not cheap. However, our algorithm is easy to parallelize and we expect that large-scale parallelization will significantly diminish this disadvantage and will make the CV analysis using SA practical.

%The remainder of this paper is organized as follows.
%In the next section, we introduce our SA algorithm.
%In \Rsec{SA}, how SA is utilized for evaluating CV is mentioned.
%A known problem on applying CV for SAP is also explained.
%In \Rsec{Result}, the performance of our approach is
%tested using synthetic data. Application results to real world data sets are also shown.
%\Rsec{Summary} is devoted to summary.

%%%%%%%%%%%%%%%%%%%%%%%%%%%%%%%%%%%%%%%%%%%%%%%%%%%%%
%%%%%%%%%%%%%%%%%%%%%%%%%%%%%%%%%%%%%%%%%%%%%%%%%%%%%
%%%%%%%%%%%%%%%%%%%%%%%%%%%%%%%%%%%%%%%%%%%%%%%%%%%%%
%\section{Model Setup}\Lsec{Formulation}
%For the purpose of performance analysis,
%we assume a synthetic model in which vector $\V{y}\in \mR^{M}$ is generated from the following linear process
%\be
%\V{y}=A\V{x}_{0}+\V{\xi},
%\Leq{y-def}
%\ee
%where $\V{\xi}\in \mR^{M}$ is a noise vector, each component of which is drawn from the zero-mean normal distribution with variance $\sigma_{\xi}^2$, $\mathcal{N}(0,\sigma_{\xi}^2)$ and
%$\V{x}_{0}$ is an planted sparse representation whose sparsity, the number of non-zero components, is assumed to be $N_{0c}<M$. Our objective is to find an appropriate sparse vector that approximates $\V{y}$,
%for which we solve \Req{SAP}. We denote the solution as $\hat{\V{x}}$.

%%%%%%%%%%%%%%%%%%%%%%%%%%%%%%%%%%%%%%%%%%%%%%%%%%%%%
%%%%%%%%%%%%%%%%%%%%%%%%%%%%%%%%%%%%%%%%%%%%%%%%%%%%%
\section{Sampling formulation and simulated annealing}\Lsec{Sampling formulation}
Let us introduce a binary vector $\V{c}=(c_i)\in \lbb 0,1\rbb^{N}$, which indicates the column vectors used to approximate $\V{y}$: If $c_i=1$, the $i$th column of $A$, $\V{a}_i$,  is used; if $c_i = 0$, it is not used. We call this binary variable {\em sparse weight}. Given $\V{c}$, the optimal coefficients, $\V{x}(\V{c})$, are expressed as
\be
\V{x}(\V{c})=\argmin_{\V{x}}||\V{y}-A(\V{c}\circ \V{x}) ||_2^2,
\Leq{x(c)}
\ee
where $(\V{c}\circ \V{x})_i=c_i x_i$ represents the Hadamard product. The components of $\V{x}(\V{c})$ for the zero components of $\V{c}$ are actually indefinite, and we set them to be zero. The corresponding RSS is thus defined by
\be
\mc{E}(\V{c}|\V{y},A)=M\epsilon(\V{c}|\V{y},A)=\frac{1}{2}||\V{y}-A\V{x}(\V{c})||_2^2.
\Leq{epsilon(c)}
\ee

To perform sampling, we employ the statistical mechanical formulation in Ref.~\cite{Obuchi:16a} as follows. By regarding $\mc{E}$ as an ``energy'' and introducing an ``inverse temperature'' $\beta$,  we can define a Boltzmann distribution as
\be
P(\V{c}|\beta;\V{y},A)=\frac{1}{G}\delta\lb \sum_{i}c_i-\nnzero \rb e^{-\beta \mc{E}(\V{c}|\V{y},A)},
\Leq{P(c)}
\ee
where $G$ is the ``partition function''
\be
G=G(\beta|\V{y},A)=\sum_{\V{c}}\delta\lb \sum_{i}c_i-\nnzero \rb e^{-\beta \mc{E}(\V{c}|\V{y},A)}.
\ee
In the limit of $\beta \to \infty$, \Req{P(c)} is guaranteed to concentrate on the solution of \Req{SAP}.
Therefore, sampling $\V{c}$ at $\beta \gg 1$ offers an approximate solution of \Req{SAP}.

Unfortunately, directly sampling $\V{c}$ from \Req{P(c)} is computationally difficult.
To resolve this difficulty, we employ a Markov chain dynamics whose equilibrium
distribution accords with \Req{P(c)}. Among many choices of such dynamics, we employ the standard
Metropolis-Hastings rule~\cite{Hasting:70}.
A noteworthy remark is that a trial move of the sparse weights, $\V{c}\to \V{c}'$, is generated by ``pair flipping'' two sparse weights, one equal to $0$ and the other equal to $1$. Namely, choosing an index $i$ of the sparse weight from ${\rm ONES}\equiv \{k|c_{k}=1 \}$ and another index $j$ from ${\rm ZEROS}\equiv \{k|c_{k}=0  \}$, we set $\V{c}'=\V{c}$, except for the counterpart of $(c_i,c_j)=(1,0)$, which is given as $(c'_i,c'_j)=(0,1)$. This flipping can keep the sparsity constant during the update.
A pseudo-code of the MC algorithm is given in \Rcode{MC} as a reference.

The most time-consuming part in the algorithm is the evaluation of $\mc{E}^\prime$ denoted in the sixth line; the naive operation for it requires $O(M\nnzero^2+\nnzero^3)$ since matrix inversion of the gram matrix $A(\V{c})^{\rm T} A(\V{c})$ is involved, where ${\rm T}$ and $A(\V{c})$ stand for the matrix transpose and the submatrix of $A$ that is composed of column vectors of $A$ whose column indices belong to ${\rm ONES}$, respectively. However, since the flip $\V{c} \to \V{c}^\prime$ changes $A(\V{c})$ only by two columns, one can reduce this computational cost to $O(M\nnzero+\nnzero^2)$ using a matrix inversion formula~\cite{Obuchi:16a}. This implies that, when the average number of flips per variable (MC steps) is kept to a constant, the computational cost of the algorithm scales as $O(MN\nnzero)$ per MC step in the dominant order since $M > \nnzero$.

In general, a longer time is required for equilibrating MC dynamics as
$\beta$ grows larger. Furthermore, the dynamics has the risk of being trapped by trivial local minima of \Req{epsilon(c)} if $\beta$ is fixed to a very large value from the beginning. A practically useful technique for avoiding these inconveniences is to start with a sufficiently small $\beta$ and gradually increase it, which is termed simulated annealing (SA)~\cite{Kirkpatrick:83}. As $\beta \to \infty$, $\V{c}$ is no longer updated, and final configuration $\V{c}_{\rm fin}$ is expected to lead to a solution that is very close (or identical) to the optimal solution in \Req{SAP},  i.e., $\hat{\V{x}}(\nnzero) \approx \V{x}(\V{c}_{\rm fin})$.

SA is mathematically guaranteed to find the globally optimal solution of \Req{SAP} if
$\beta$ is increased to infinity slowly enough in such a way that $\beta(t) < C \log (t+2)$ is satisfied,
where $t$ is the counter of the MC dynamics and $C$ is a time-independent constant~\cite{Geman:84}.
Of course, this schedule is practically meaningless, and a much faster schedule is employed
generally. In Ref.~\cite{Obuchi:16a}, we examined the performance of SA with a very
rapid annealing schedule for a synthetic model whose properties of fixed $\beta$ can be analytically evaluated.
Comparison between the analytical and the experimental results
indicates that the rapid SA performs quite well unless a phase transition of a certain type
occurs at relatively low $\beta$. Owing to the analysis of the synthetic model, the range of system parameters
in which the phase transition occurs is rather limited.
We therefore expect that SA serves as a promising approximate solver for
\Req{SAP}.

%Motivated by this earlier study,
%we here again use SA for evaluating cross validation (CV) in the sampling approach. Our SA algorithm is based on the standard MC method with the Metropolis criterion as~\cite{Obuchi:16a}. A noteworthy remark is that a trial move of sparse weights, $\V{c}\to \V{c}'$, is generated by ``pair flipping'' two sparse weights, one equal to $0$ and the other equal to $1$. Namely, choosing an index $i$ of the sparse weight from ${\rm ONES}\equiv \{k|c_{k}=1 \}$ and another index $j$ from ${\rm ZEROS}\equiv \{k|c_{k}=0  \}$, we set $\V{c}'=\V{c}$, except for the counterpart of $(c_i,c_j)=(1,0)$, which is given as $(c'_i,c'_j)=(0,1)$. This flipping keeps the sparsity constant during the update. A pseudo-code of the MC algorithm is given in \Rcode{MC} as a reference.
%%%%%%%%%%%%%%%%%%%%%%%%%%%%%%%
\alglanguage{pseudocode}
\begin{algorithm}[htbp]
\caption{MC update with pair flipping}\Lcode{MC}
\begin{algorithmic}[1]
\Procedure{MCpf}{$\V{c},\beta,\V{y},A$}\Comment{MC routine}
	\State ${\rm ONES} \lA \{k|c_k=1\},~{\rm ZEROS} \lA \{k|c_k=0\}$
	\State randomly choose $i$ from ONES and $j$ from ZEROS
	\State $\V{c}' \lA \V{c}$
	\State $(c'_i, c'_j) \lA (0,1)$
	\State $(\mc{E},\mc{E}')\lA (\mc{E}(\V{c}|\V{y},A),\mc{E}(\V{c}'|\V{y},A))$
	\State $p_{\rm accept} \lA \max(1,e^{-\beta\lb \mc{E}' -\mc{E}\rb } )$
	\State generate a random number $r\in [ 0,1]$
	\If {$ r < p_{\rm accept} $}
		\State $\V{c} \lA \V{c}'$
	\EndIf
	\State \Return $\V{c}$
\EndProcedure
\end{algorithmic}
\end{algorithm}
%%%%%%%%%%%%%%%%%%%%%%%%%%%%%%%
%
%Using this MC routine, our SA algorithm updates the configuration of $\V{c}$ while gradually decreasing the temperature $T=1/\beta\gg 1$.

%%%%%%%%%%%%%%%%%%%%%%%%%%%%%%%%%%%%%%%%%%%%%%%%%%%%%
%%%%%%%%%%%%%%%%%%%%%%%%%%%%%%%%%%%%%%%%%%%%%%%%%%%%%
%%%%%%%%%%%%%%%%%%%%%%%%%%%%%%%%%%%%%%%%%%%%%%%%%%%%%
%\section{CV and its theoretical background}\Lsec{SA}
\section{Employment of SA for cross validation}\Lsec{SA}
CV is a framework designed to evaluate the generalization ability of statistical models/learning systems
based on a given set of data. In particular, we examine the leave-one-out (LOO) CV, 
but its generalization to the $k$-fold CV is straightforward.

%%%%%%
In accordance with the cost function of \Req{SAP}, we define the generalization error
\be
\epsilon_{\rm g} =\frac{1}{2} \overline{ \left (y_{M+1} -\sum_{i=1}^N A_{(M+1),i} x_i\right )^2}
\Leq{generalization}
\ee
as a natural measure for evaluating the generalization ability of $\V{x}$,
where $\overline{\cdots}$ denotes the expectation with respect to the ``unobserved'' $(M+1)$th data
$\left (\{A_{(M+1),i} \}, y_{M+1} \right )$.
LOO CV assesses the LOO CV error (LOOE)
\be
\LOOE(\nnzero|\V{y},A) =\frac{1}{2M}\sum_{\mu=1}^{M}\lb y_{\mu}-\sum_{i=1}^N
A_{\mu i}x^{\bs \mu}_{i}(\V{c}^{\bs \mu})\rb^2
\Leq{LOOE}
\ee
as an estimator of \Req{generalization} for the solution of \Req{SAP},
where $\V{x}^{\bs \mu}(\V{c}^{\bs \mu}) =(x^{\bs \mu}_{i}(\V{c}^{\bs \mu}))$ is the solution of \Req{SAP}
for the ``$\mu$th LOO system,'' which is defined by removing the $\mu$th data $\left (\{A_{\mu i}\}, y_\mu \right )$
from the original system.
One can evaluate \Req{LOOE}  by independently applying SA to each of the $M$ LOO systems.

The LOOE of \Req{LOOE} depends on $\nnzero$ through $\V{c}^{\bs \mu}$, and hence we can determine its ``optimal'' value from the minimum of $\LOOE(\nnzero|\V{y},A)$ by sweeping $\nnzero$. Compared to the case of a single run of SA at a given $\nnzero$, the computational cost for LOO CV is increased by a certain factor. This factor is roughly evaluated as $O(M\times | \nnzeroset | \times \nnzeromax )$ when varying $\nnzero$ in the set of $\nnzeroset$ in which the maximum value is $\nnzeromax$. However, this part of the computation can be easily parallelized and, therefore, may not be so problematic when sufficient quantities of CPUs and memories are available. In the next section, the rationality of the proposed approach is examined by application to a synthetic model and a real-world data set.

Before closing this section, we want to remark on 
two important issues.
%an important issue. 
For this, we assume that $\V{y}$ is generated by a true sparse vector
$\V{x}_0$ as
\be
\V{y}=A\V{x}_{0}+\V{\xi},
\Leq{y-def}
\ee
where $\V{\xi}\in \mR^{M}$ is a noise vector whose entries are uncorrelated with one another.

The first issue is about the accuracy in inferring $\V{x}_0$. 
When $A$ is provided as a column-wisely normalized zero mean 
random matrix whose entries are uncorrelated with one another, \Req{generalization} 
is linearly related to the squared distance between the true and inferred vectors, 
$\V{x}_0$ and $\hat{\V{x}}$, as
\be
\epsilon_{\rm g} = \frac{d_1}{N}||\hat{\V{x}}-\V{x}_0||_2^2+d_0
\Leq{MSE}
\ee
\cite{Seung:92,Opper:96,Obuchi:16b},
where $d_1$ and $d_0$ are positive constants. 
%$\V{x}_0$ and $\hat{\V{x}}$, which is given as 
%\be
%d^2 \! \left (\hat{\V{x}}, \V{x}_0 \right )=\frac{1}{N}||\hat{\V{x}}-\V{x}_0||_2^2
%\Leq{MSE}
%\ee
%\cite{Seung:92,Opper:96,Obuchi:16b}. 
%choosing the minimum of 
 Hence, minimizing the estimator of \Req{generalization}, i.e. \Req{LOOE}, 
%has a meaning of 
leads to %optimizing the accuracy of $\hat{\V{x}}$ in the sense of %\Req{MSE}. 
minimizing the squared distance from the true vector $\V{x}_0$. 
The same conclusion has also been obtained for %the $\ell_1$-regularization approach 
LASSO in %the asymptotic theory taking 
the limit of $M\to \infty$  while keeping $N$ and $\nnzero$ finite~\cite{Homrighausen:14}, 
where it is not needed to assume absence of correlations in $A$. 
%The correspondence between the generalization error and \Req{MSE}
% also justifies minimization of \Req{LOOE} in the context of the recovery of the sparse representation $\V{x}_0$. 

The second issue is about the difficulty in  
%For a wide class of linear models, 
identifying sparse weight vector $\V{c}_0$ of $\V{x}_0$ using CV, 
which is known to fail even when $M/N \to \infty$~\cite{Shao:93}. Actually, we have tried a naive approach to identify $\V{c}_0$  by directly minimizing a CV error with the use of SA and confirmed that it does not work. A key quantity for this is another type of LOOE:
\be
\LOOEtilde(\V{c}|\V{y},A)=\frac{1}{2M}\sum_{\mu=1}^{M}\lb y_{\mu}-\sum_{i=1}^N A_{\mu i}x^{\bs \mu}_{i}(\V{c}) \rb^2.
\Leq{LOOE2}
\ee
This looks like \Req{LOOE}, but is different in that $\V{c}$ is common among all the terms.
%Therefore, one can select a certain $\V{c}$ by minimization of $\LOOEtilde(\V{c}|\V{y},A)$, which can be performed by a single run of SA handling $\LOOEtilde(\V{c}|\V{y},A)$ as energy function of $\V{c}$.
It may be natural to expect that the sparse weight minimizing \Req{LOOE2}, $\tilde{\V{c}}=\mathop{\rm argmin}_{\V{c}}\{\LOOEtilde(\V{c}|\V{y},A)\}$, is the ``best'' $\V{c}$ that converges to $\V{c}_0$ in the limit of $M/N \to \infty$. Unfortunately, this is not true;  in fact, $\LOOEtilde(\V{c}|\V{y},A)$ in general tends to decrease as $\nnzero$ increases, irrespective of the value of $||\V{x}_0||_0$~\cite{Shao:93}. We have confirmed this by conducting SA, handling $\LOOEtilde(\V{c}|\V{y},A)$ as an energy function of $\V{c}$. Therefore, minimizing \Req{LOOE2} can neither identify $\V{c}_0$ nor offer any clue for determining $\nnzero$.

We emphasize that minimization of \Req{LOOE} can be utilized to determine $\nnzero$ to optimize the generalization ability, although it does not have the ability to identify $\V{c}_0$ either. The difference between these two issues is critical and confusing, as several earlier studies have provided some controversial implications to the usage of CV in sparse inference on 
linear models~\cite{Bousquet:02,Leng:06,Shalev-Shwartz:09,Xu:12,Homrighausen:14}.

%The second issue is about justification of using \Req{LOOE} for
%the hyper parameter determination.
%Under the current assumption, $\LOOE$ plays a role of an estimator of the prediction error for
%an ``unobserved $M+1$th'' data.
%The second issue is about the consistency of the inference of $\V{x}_0$. 
%Each term in \Req{LOOE} is linearly related to
%the squared distance between the true and inferred vectors, 
%$\V{x}_0$ and $\hat{\V{x}}$, 
%\be
%\mc{M}(\hat{\V{x}})=\frac{1}{N}||\hat{\V{x}}-\V{x}_0||_2^2,
%\Leq{MSE}
%\ee
%if each component in $A$ and each component of the noise vector
%are uncorrelated with one another. 
%%in the limit $N,M\to \infty$ while the aspect ratio $\alpha=M/N$ is kept finite~
%\cite{Seung:92,Opper:96,Obuchi:16b}. Hence, choosing
%the minimum of LOOE also has a meaning of finding the vector
%that is most close to $\V{x}_0$ in the sense of \Req{MSE}. The same conclusion has been obtained for the $\ell_1$-regularization approach in the asymptotic theory taking the $M\to \infty$ limit while keeping $N$ and $\nnzero$ finite~\cite{Homrighausen:14}. In the asymptotic theory, it is not needed to assume absence of correlations in $A$. Anyway, this correspondence of the CV error with \Req{MSE}
%justifies the use of it in the context of the recovery of the sparse representation.

%%%%%%%%%%%%%%%%%%%%%%%%%%%%%%%%%%%%%%%%%%%%%%%%%%%%%
%%%%%%%%%%%%%%%%%%%%%%%%%%%%%%%%%%%%%%%%%%%%%%%%%%%%%
%%%%%%%%%%%%%%%%%%%%%%%%%%%%%%%%%%%%%%%%%%%%%%%%%%%%%
\section{Result}\Lsec{Result}

%%%%%%%%%%%%%%%%%%%%%%%%%%%%%%%%%%%%%%%%%%%%%%%%%%%%%
%%%%%%%%%%%%%%%%%%%%%%%%%%%%%%%%%%%%%%%%%%%%%%%%%%%%%
\subsection{Test on a synthetic model}\Lsec{Test on}
%%%%%%%%%%%%%%%%%%%%%
\begin{figure}[t]
\begin{center}
\includegraphics[width=\columnwidth,height=0.66\columnwidth]{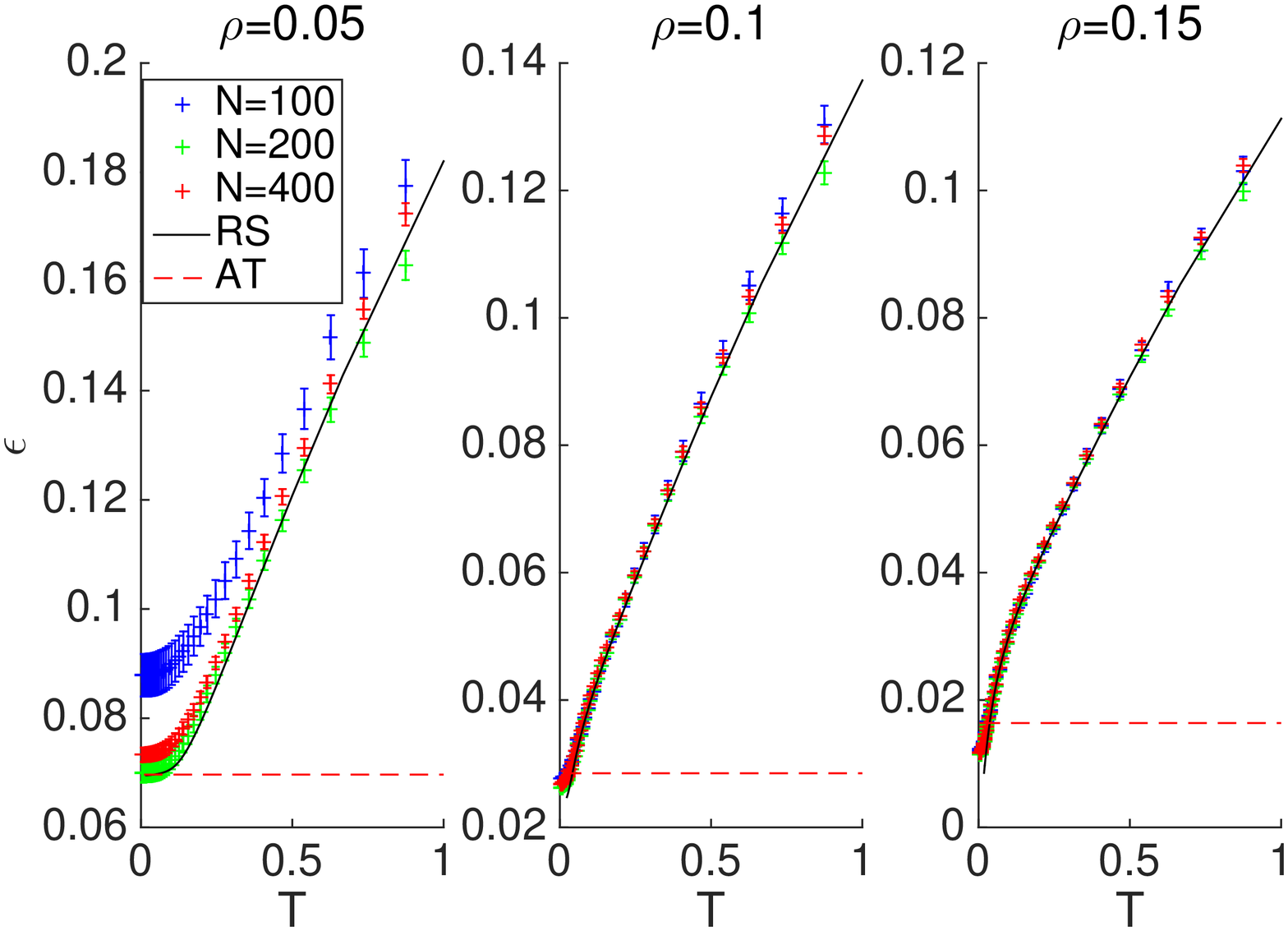}
\end{center}
\caption{RSS per component $\epsilon$ versus $T=\beta^{-1}$ observed in the annealing process for $N=100$, $200$, and $400$.
Curves represent the RS predictions for \Req{P(c)}. The replica symmetry is broken owing to the AT instability
below the broken lines.
}
\Lfig{beta_vs_Err}
\end{figure}
%%%%%%%%%%%%%%%%%%%%%
%%%%%%%%%%%%%%%%%%%%%
\begin{figure}[t]
\begin{center}
\includegraphics[width=0.9\columnwidth,height=0.68\columnwidth]{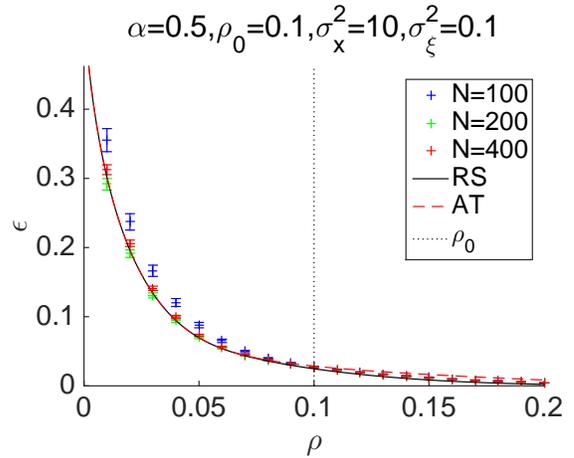}
\end{center}
\caption{ RSS per component $\epsilon$ finally achieved by SA (symbols), its RS assessments
for $\beta \to \infty$ (full curve) and at the onset of the AT instability (red broken curve), plotted against $\rho=\nnzero/N$.
}
\Lfig{ERR_vs_rho}
\end{figure}
%%%%%%%%%%%%%%%%%%%%%
%%%%%%%%%%%%%%%%%%%%%
\begin{figure}[t]
\begin{center}
\includegraphics[width=0.9\columnwidth,height=0.68\columnwidth]{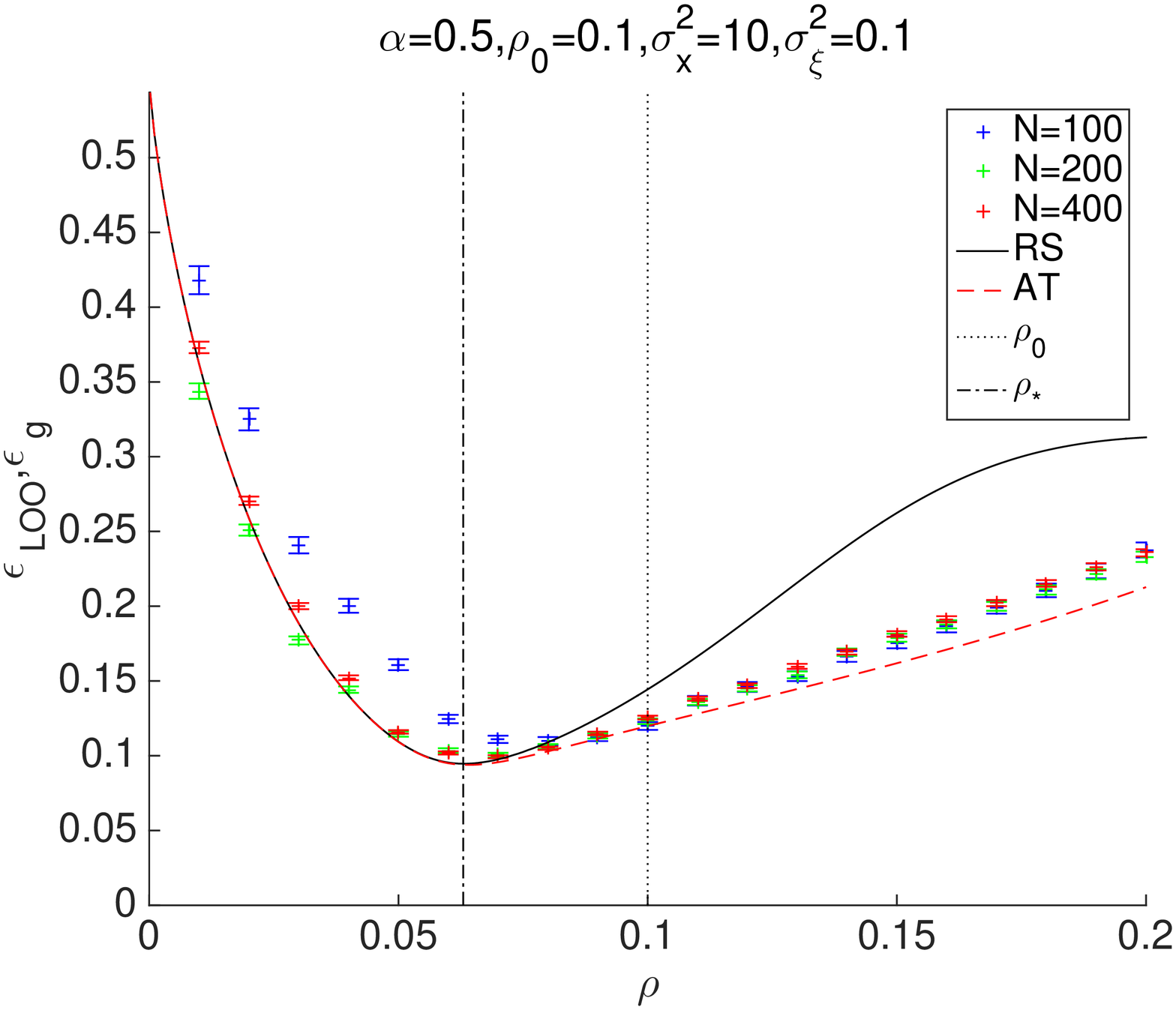}
\end{center}
\caption{ $\LOOE$ evaluated by solutions of SA (symbols) and RS assessments of typical $\epsilon_{\rm g}$ for
$\beta \to \infty$ (full curve) and at the onset of the AT instability (red broken curve), plotted against $\rho=\nnzero/N$.
% by SA (symbols), RS assessment for $\beta \to \infty$ (full curve), and the value at the onset of the AT instability (red broken curve).
%The minimum points of full and broken curves, $\rho_*$ and $\rho_*^{\rm AT}$, are almost overlapped in this figure.
}
\Lfig{ERRmin_LOOE}
\end{figure}
%%%%%%%%%%%%%%%%%%%%%
%%%%%%%%%%%%%%%%%%%%%

We first examine the utility of our methodology by applying it to a synthetic model in which vector $\V{y}$ is generated in the manner of \Req{y-def}.
For analytical tractability, we assume that $A$ is a simple random matrix whose entries are
independently sampled from ${\cal N}(0, N^{-1})$ and that each component of $\V{x}_0$ and $\V{\xi}$ is also independently generated from $(1-\rho_0)\delta(x)+\rho_0 {\cal N}(0, \sigma_x^2)$ and ${\cal N}(0, \sigma_\xi^2)$, respectively. Under these assumptions, an analytical technique based on the replica method of statistical mechanics makes it possible to theoretically assess the typical values of various macroscopic quantities when $\V{c}$ is generated from \Req{P(c)} as $N \to \infty$ keeping $\alpha =M/N $ %finite~\cite{Nakanishi:15,Obuchi:15}. 
finite~\cite{Nakanishi:15}. 
We performed a theoretical assessment under the so-called {\em replica symmetric (RS)} assumption.

In the experiment, system parameters were fixed as $\rho_0=0.1$, $\alpha=0.5$, $\sigma_x^2=10$, and $\sigma_\xi^2=0.1$. The annealing schedule was set as
\be
\beta_a=\beta_0+r^{a-1}-1,~\tau_a=\tau,~(a=1,\cdots,100),
\Leq{schedule}
\ee
where $\tau_a$ denotes the typical number of MC flips per component for a given value of inverse temperature $\beta_a$. We set $\tau=5$, $\beta_0=10^{-8}$, and $r=1.1$ as default parameter values.
Thus, the maximum value of $\beta$ was  $\beta_{100}\approx 1.3 \times 10^{4}$. The examined system sizes were $N=100$, $200$, and $400$. We took the average over $N_{\rm samp}=100$ different samples. %for $N=100$ and $200$ and over  $N_{\rm samp}=40$ samples for $N=400$. 
The error bar is given by the standard deviation among those samples divided by $\sqrt{N_{\rm samp}-1}$.

\Rfig{beta_vs_Err} shows how RSS per component $\epsilon$ in \Req{SAP}
depends on $T=\beta^{-1}$ during the annealing process for $\nnzero/N\equiv \rho=0.05$, $0.1$, and $0.15$.
The data of SA (symbols) for $N=100$, $200$, and $400$ totally exhibit a considerably good accordance with the theoretical prediction (curves) for \Req{P(c)} despite the very rapid annealing schedule of \Req{schedule}, in which $c_i$ $(i=1,2,\ldots,N)$ is flipped only five times on average at each value of
$\beta=\beta_a$. The replica analysis indicates that the {\em replica symmetry breaking (RSB) }
due to the {\em de Almeida-Thouless (AT) instability}~\cite{AT:78} occurs below the broken lines.
However, the RS predictions for $\beta \to \infty$ still serve as the lower bounds of $\epsilon_{\rm min}$ 
even in such cases~\cite{Obuchi:10,Obuchi:15b}. %The data show
\Rfig{ERR_vs_rho} shows
 that the values achieved by SA are fairly close to the lower bounds, 
 implying that SA can find nearly optimal solutions of \Req{SAP}.

Let us denote \Req{generalization} of typical samples generated from \Req{P(c)}
at inverse temperature $\beta$ as $\epsilon_{\rm g}(\beta)$.
The generalization error of the optimal solution of \Req{SAP} is assessed as $\epsilon_{\rm g}(\infty)$.
\Rfig{ERRmin_LOOE} %is the plot of
plots %LOOE \Req{LOOE}
the $\LOOE$ evaluated by the solutions of SA (symbols) and the RS evaluations of %generalization error \Req{generalization}
$\epsilon_{\rm g}(\infty)$ (full curve) and $\epsilon_{\rm g}(\beta_{\rm AT})$
(red broken curve), against $\rho=\nnzero/N$.
Here, $\beta_{\rm AT}$ is the critical inverse temperature at which RSB occurs owing to the AT instability.
The three plots accord fairly well with one another
in the left of their minimum point $\rho_* \sim 0.063$, whereas there are considerable discrepancies between
$\epsilon_{\rm g}(\infty)$ and the other two plots for $\rho > \rho_*$.

The discrepancies are considered to be caused by RSB.
For $\beta > \beta_{\rm AT}$, the MC dynamics tends to be trapped by a metastable state.
This makes it difficult for SA to find the global minimum of 
$\epsilon(\V{c})$, 
%$\epsilon(\V{c})$~\cite{Krzakala:07},
which explains why the SA's results are close to $\epsilon_{\rm g}(\beta_{\rm AT})$.
Fortunately, this trapping works beneficially for the present purpose of
%finding $\nnzero$ that yields higher generalization ability
raising the generalization ability by lowering
$\epsilon_{\rm g}$, as seen in \Rfig{ERRmin_LOOE}. As far as we have examined, for fixed $\rho$,
$\epsilon_{\rm g}(\beta_{\rm AT})$  never exceeds the RS evaluation of $\epsilon_{\rm g}(\infty)$
and is always close to $\LOOE$ of SA's results. These imply that the generalization ability achieved by tuning $\nnzero=N\rho$
using the SA-based CV is no worse than that obtained when CV is performed by exactly solving \Req{SAP} for LOO systems. This is presumably because, for a large $\rho$, the optimal solution of \Req{SAP} overfits the observed (training) data and its generalization ability becomes worse than that of $\V{x}(\V{c})$ typically sampled at appropriate values of $\beta(<\infty)$.

\subsection{Application to a real-world data set}\Lsec{Application to}

\begin{table}
\small
\begin{center}
\begin{tabular}{l ||c|c|c|c |c }
$\nnzero$ & 1 & 2 & 3 & 4 & 5 \\
\hline
$\LOOE$ & 0.0328& 0.0239 &  0.0281 & 0.0331 & 0.0334 \\
\end{tabular}
\caption{\label{tab:supernova_LOOE}
LOO CV error obtained for $\nnzero=1$--$5$ for the type Ia supernova data set. }
\end{center}
\vspace*{-0.3cm}
\end{table}

%%%%%%%%%%%%%%%%%%%%%%%%%%%%%%%%%%%%%%

\begin{table}
\small
\begin{center}
\begin{tabular}{ l || c | c | c | c | c}
\multicolumn{6}{c}{$\nnzero=1$} \\ \hline
variable & 2 & $*$& $*$ & $*$ & $*$ \\
\hline
times selected & 78 & 0 & 0 & 0 & 0\\
\end{tabular}
\vspace*{0.2cm}
%%%
\begin{tabular}{ l || c | c | c | c | c}
\multicolumn{6}{c}{$\nnzero=2$} \\ \hline
variable & 2 & 1 & 275 & $*$ & $*$ \\
\hline
times selected & 78 & 77 & 1 & 0 & 0\\
\end{tabular}
\vspace*{0.2cm}
%%%
\begin{tabular}{ l || c | c | c | c | c}
\multicolumn{6}{c}{$\nnzero=3$} \\ \hline
%variable & 2 & 1 & 14 & 15 & 5 \\
variable & 2 & 1 & 233 & 14 & 69 \\
\hline
%times selected & 78 & 75 & 62 & 8 & 2\\
times selected & 78 & 76 & 69 & 3 & 2\\
\end{tabular}
\vspace*{0.2cm}
%%%
\begin{tabular}{ l || c | c | c | c | c}
\multicolumn{6}{c}{$\nnzero=4$} \\ \hline
%variable & 2 & 1 & 233 & 96 & 225 \\
variable & 2 & 1 & 233 & 94 & 225 \\
\hline
%times selected & 78 & 49 & 36 & 31 & 14\\
times selected & 78 & 59 & 56 & 49 & 13\\
\end{tabular}
\vspace*{0.2cm}
%%%
\begin{tabular}{ l || c | c | c | c | c}
\multicolumn{6}{c}{$\nnzero=5$} \\ \hline
%variable & 2 & 170 & 225 & 6 & 233 \\
variable & 2 & 36 & 223 & 225 & 6 \\
\hline
%times selected & 78 & 39 & 38 & 35 & 32 \\
times selected & 78 & 37 & 33 & 31 & 27 \\
\end{tabular}
\caption{
\label{tab:supernova_count}
The top five variables selected by the $M=78$ LOO CV
for $\nnzero=1$--$5$.
}
\end{center}
\vspace*{-0.3cm}
\end{table}

We also applied our SA-based analysis to a data set from the SuperNova DataBase
provided by the Berkeley Supernova Ia program \cite{Berkeley,Silverman:12}.
%%%
Screening based on a certain criteria 
yields a reduced data set of $M = 78$ and $N = 276$ \cite{Uemura:15}.
%%%
%The system sizes of this data set are $M = 78$ and $N = 276$.
The purpose of the data analysis is to select a set of explanatory variables
relevant for predicting the absolute magnitude at the maximum of type Ia supernovae by linear regression. 
%from the reduced data set. 
%%%

Following a conventional treatment of linear regression,
we preprocessed both the absolute magnitude at the maximum (dependent variable) and the $276$ candidates of explanatory variables to have zero means.
We performed the SA for $M=78$ LOO systems of the preprocessed data set. The result of 
one 
single experiment with varying $\nnzero$ is given in Tables \ref{tab:supernova_LOOE} and \ref{tab:supernova_count}. Table \ref{tab:supernova_LOOE} provides the values of LOOE, which shows that $\LOOE$ is minimized at $\nnzero=2$.

Possible statistical correlations between explanatory variables,
which were not taken into account in the synthetic model in \Rsec{Test on},
could affect the results of linear regression~\cite{Belsley:91}.
The CV analysis also offers a useful clue for checking this risk.
Examining the SA results of $M(=78)$ LOO systems, we could count
how many times each explanatory variable was selected,
which could be used for evaluating the reliability of the variable~\cite{Meinshausen:10}.
{Table \ref{tab:supernova_count} summarizes the results for five variables from the top 
for $\nnzero=1$--$5$.}
This indicates that no variables other than ``2,'' which stands for {\em color},
were chosen stably, whereas variable ``1,'' representing {\em light curve width},
was selected with high frequencies for $\nnzero \le 3$.
Table \ref{tab:supernova_count} shows that the frequency of ``1'' being selected is significantly reduced for $\nnzero \geq 4$. 
These are presumably due to the strong statistical correlations between ``1'' and the newly added variables, suggesting the low reliability of the CV results for $\nnzero \ge 4$. 
In addition, {for $\nnzero \geq 4$, we observed that 
the results varied depending on samples generated by the MC dynamics in SA, which}
implies that there exist many local minima in \Req{epsilon(c)} of LOO systems for $\nnzero \geq 4$. These observations mean that we could select at most only {\em color} and {\em light curve width}  as the explanatory variables relevant for the absolute magnitude prediction with a certain confidence.
This conclusion is consistent with that of %a recent paper 
\cite{Uemura:15},
in which the relevant variables were selected by LASSO,
combined with hyper parameter determination following the {\em ad hoc} ``one-standard-error rule,''
and with the comparison between several resulting models.

%%%%%%%%%%%%%%%%%%%%%%%%%%%%%%%%%%%%%%%%%%%%%%%%%%%%%
%%%%%%%%%%%%%%%%%%%%%%%%%%%%%%%%%%%%%%%%%%%%%%%%%%%%%
%%%%%%%%%%%%%%%%%%%%%%%%%%%%%%%%%%%%%%%%%%%%%%%%%%%%%
\section{Summary}\Lsec{Summary}
We examined the abilities and limitations of simulated annealing (SA) for sparse approximation problem,
in particular, when employed for determining degrees of freedom by cross validation (CV).
Application to a synthetic model indicates that SA can find nearly optimal solutions for \Req{SAP},
and when combined with the CV framework, it can optimize the generalization ability.
Its utility was also tested by application to a real-world supernova data set.

Although we focused on the use of SA, samples at finite temperatures contain
useful information for SAP. How to utilize such information is currently under investigation.

\section*{Acknowledgments}
This work was supported by JSPS KAKENHI under grant numbers 26870185 (TO) and 25120013 (YK).
The UC Berkeley SNDB is acknowledged for permission of using the data set of \Rsec{Application to}.
Useful discussions with Makoto Uemura and Shiro Ikeda on the analysis of the data set
are also appreciated.


\end{document}